\documentclass[twocolumn,preprintnumbers,aps,showpacs,amsmath,amssymb]{revtex4}
\usepackage{graphicx}
\usepackage{bm}
\usepackage{float}

\newcommand{\mL}{{\mathcal{L}}}
\newcommand{\mE}{{\mathcal{E}}}
\newcommand{\mB}{{\mathcal{B}}}

\newcommand{\csch}{\mathrm{csch}}

\begin{document}

\preprint{v2 of arXiv:1805.03622 [hep-ph]}

\title{Vacuum stabilized by anomalous magnetic moment}

\author{Stefan Evans}
\author{Johann Rafelski}
\affiliation{Department of Physics, The University of Arizona, Tucson, AZ 85721, USA}
\date{June 9, 2018}

\begin{abstract}
An analytical result for Euler-Heisenberg effective action, valid for electron spin $g-$factor $|g|<2$, was extended to the domain $|g|>2$ via discovered periodicity of the effective action. This allows for a simplified computation of vacuum instability modified by the electron's measured $g=2.002319$. We find a strong suppression of vacuum decay into electron positron pairs when magnetic fields are dominant. The result is reminiscent of mass catalysis by magnetic fields.
\end{abstract}



\maketitle

\section{Introduction}

We explore the effect of anomalous electron spin $g$-factor on vacuum instability in Euler Heisenberg (EH) effective action~\cite{Heisenberg:1935qt,Weisskopf:1996bu,Schwinger:1951nm}. In the wake of theoretical development of EH action, studies of the nonlinear QED vacuum have almost always relied on an electron spin $g$-factor of exactly 2. Since both real and imaginary parts of the effective action are modified by an anomalous $g\ne 2$-factor, the rate of vacuum decay in strong fields by pair production is affected. We study this effect here.

Employing an anomalous $g\ne 2$-factor is an improvement on the EH effective theory in its current form, serving as an introduction of higher order corrections into the theory. EH action has been extended to $|g|<2$~\cite{OConnell:1968spc,Dittrich:1977ee,Kruglov:2001dp,Labun:2012jf}. A finite regularized expression for all $g$ was obtained by exploiting periodicity of the action~\cite{Rafelski:2012ui}. While a magnetic field added to an electric field enhances the vacuum decay rate at $g=2$~\cite{Cho:2000ei,Labun:2008re}, we will show that for strong magnetic fields a significant suppression to particle production arises for $g\neq 2$, an effect reminiscent of magnetic mass catalysis~\cite{Gusynin:1995gt,Gusynin:1999ti,Ferrer:2000ed,Elizalde:2002ca,Shovkovy:2012zn} in that it can be rather precisely described by modification of particle mass induced by the magnetic field.

Strong magnetic fields are found on surfaces of magnetized neutron stars (magnetars), suggested to possess $\mB$ fields $\sim 10^2$ above EH critical field ($\mE_\mathrm{EH}=m^2/e$ in Lorentz-Heaviside natural units, where $m$ is electron mass)~\cite{Duncan:1992hi,Kouveliotou:1998ze,Harding:2006qn}. Nonlinear QED phenomena in such scenarios have been studied extensively, see~\cite{Ruffini:2009hg,Kim:2011hr} and references therein.
Even stronger $\mB$ fields are produced for ultra-short time intervals in relativistic heavy ion collisions~\cite{Rumrich:1991xs,Baur:2007fv,Ruffini:2009hg,Rafelski:2016ixr}.

The here presented results are a step towards understanding of non-perturbative QED vacuum structure in ultra-strong fields at the scale $\mE_\mathrm{EH}/(g/2-1)$. Here we accommodate the non-perturbative character of modifications arising from $g\neq 2$ explicitly and quantify how for a given value of $g\neq 2$ modifications of the instability of the EH effective action arise in presence of external fields. In order to complete QED vacuum structure study in \lq\lq two loop\rq\rq\ order: second order in $\alpha\simeq 1/137$ but to all orders in external (constant) fields, it is necessary to consider i) magnetic field dependent $g $, along with ii) direct self-energy modifications of the particle mass $m$ by (constant) external fields~\cite{Jancovici:1970ep,Newton:1971pq,Constantinescu:1972qe,Tsai:1974id}. Our results presented in this work allow to insert the field dependence $g\to g_f(a,b),\, m\to m_f(a,b)$ (subscript $f$ reminds of field dependence) to obtain from the one loop EH effective action the corresponding two loop result accounting in our approach for the non-perturbative behavior of EH effective action as a function of $g$.

\section{Modified Form of Effective Action}

\subsection{$|g|<2$}

The Schwinger proper time method at $g=2$ was extended to $|g|<2$ and produces effective action~\cite{Schwinger:1951nm,Labun:2012jf}
\begin{align}
\label{eq1}
\mL_\mathrm{EH}=&\,\frac1{8\pi^2}\int_0^\infty \frac{ds}{s^3}e^{-im^2s}
 \\[0.2cm] \nonumber
\hphantom=&
\times
\left( \frac{eas\cosh[\frac g2eas]}{\sinh[eas]}\frac{ebs\cos[\frac g2ebs]}{\sin[ebs]} -1 \right)
\;,
\end{align}
where
\begin{equation}
a^2-b^2=\mE^2-\mB^2=2S\;,\;
ab=\mE\cdot\mB=P
\;
\end{equation}
and $\alpha=e^2/4\pi$ (as used in~\cite{Schwinger:1951nm}). For a pure electric field, meromorphic expansion and regularization produces temperature representation~\cite{Muller:1977mm,Labun:2012jf}
\begin{equation}
\label{meromorphic}
\mL_\mathrm{EH}=
\frac{m^2T}{16\pi^2}\int_0^\infty dE\ln[E^2-m_-^2]
\sum_\pm\ln[1+e^{\pm i\pi\frac g2}e^{-E/T}]
\;,
\end{equation}
where temperature parameter
\begin{equation}
T=\frac{e\mE}{m\pi}
\;,
\end{equation}
and mass
\begin{equation}
m_-^2=m^2-i\epsilon
\;
\end{equation}
to offset poles.
These two representations of EH action are only valid for $|g|<2$: Eq.\,(\ref{eq1}) diverges if $|g|>2$, readily shown when writing $\cosh$ and $\cos$ terms as exponentials, and meromorphic expansion in Eq.\,(\ref{meromorphic}) fails for $|g|>2$.

\subsection{$|g|>2$}

To extend to $|g|>2$, we take advantage of EH action being periodic in $g$. We repeat in abbreviated format the arguments and derivations seen in~\cite{Rafelski:2012ui}. For a constant $\mB$ field pointing in $\hat z$ the Landau orbit spectrum is obtained using the so called Klein-Gordon-Pauli generalization of the Dirac equation 
\begin{equation}
\label{Landau}
E_n=\pm\sqrt{m^2+p_z^2+Q|e\mB|[(2n+1)\mp g/2]}
\;,
\end{equation}
where $Q=\pm1$. Summing orbit quantum number $n$, we see that a shift
\begin{equation}
g\to g+4k
\;
\end{equation}
corresponds to a shift in $n$, leaving the summed states (Casimir energy,~\cite{Weisskopf:1996bu}) unchanged. Noting this periodicity and using the usual method of discrete summation~\cite{Heisenberg:1935qt,Weisskopf:1996bu} to obtain an integral representation, a periodic function introduces the Bernoulli polynomials~\cite{Euler,Apostol}. With this periodic extension of the Bernoulli polynomials, a modified meromorphic expansion is used (Eq.\,(9) in~\cite{Rafelski:2012ui}), agreeing in limit $g\to2$ with~\cite{Muller:1977mm,Cho:2000ei,Dunne:2004nc}.

At magnetic field strengths beyond 
\begin{equation}
\label{eq8}
|e\mB|<\frac{m^2}{|g/2-1|}\sim 862 m^2\;,\qquad g=g_f(0)=2.002319
\;,
\end{equation}
the energies in Eq.\,(\ref{Landau}) become imaginary, that is the associated eigenstates disappear from the spectrum and self-adjointness is lost. Since magnetic fields cannot do work, a compensating modification of mass may stabilize the vacuum in presence of ultra strong magnetic fields. We postpone this question to future study and limit the present discussion by condition Eq.\,(\ref{eq8}).

We exploit the $g$-periodicity, which applies even for $B$-dependent anomalous moment, to write 
\begin{equation}
\mL_\mathrm{EH}(g_k)=\mL_\mathrm{EH}(g_{k-1})
\;,
\end{equation}
for any real integer $k$ where
\begin{equation}
-2+4k<g_k<2+4k
\;,
\end{equation}
with cusps at the boundaries. Thus the electron anomalous moment, slightly greater than 2 (within domain $g_1$), can be transformed into the domain $|g_0|<2$ by 
\begin{equation}
\mL_\mathrm{EH}(g=2+0.002319)=\mL_\mathrm{EH}(g=-2+0.002319)
\;.
\end{equation}
We can now write the effective action for all values of $g_k$ as
\begin{align}
\mL_\mathrm{EH}=&\,
\frac1{8\pi^2}\int_0^\infty \frac{ds}{s^3}e^{-im^2s}
 \\[0.2cm] \nonumber
\hphantom=&
\!\!\!\!\!\!\!\!\!
\times
\left( \frac{eas\cosh[\frac {g_k-4k}2eas]}{\sinh[eas]}\frac{ebs\cos[\frac {g_k-4k}2ebs]}{\sin[ebs]} -1 \right)
,
\end{align}
avoiding any divergences encountered if $|g_k|>2$ appeared alone in the arguments of $\cos$ and $\cosh$. In addition, meromorphic expansion can now be applied to obtain temperature representation for any $g$: Eq.\,(\ref{meromorphic}) is also generalized to arbitrary $g_k$. 
Whenever using in the EH action an anomalous magnetic moment, of any magnitude, and any $B$-dependence, the periodic reset is implied according to
\begin{equation}
\label{g2}
\frac g2\to\frac g2-2k
\;,
\end{equation}
so that the value of $g$ is always in the principal domain $-2<g<2$. A \lq\lq raw\rq\rq\ value $|g|>2$ may never be used in Eq.\,(\ref{eq1}) as the EH effective action is ill defined in that case~\cite{Rafelski:2012ui}. Eq.\,(\ref{g2}) explains how any value of $g$ (including $g(B)$) has to be periodically reset, and only the reset value is to be used in all expressions involving the EH effective action.

\section{Stabilization of the vacuum}

\subsection{Vacuum decay in a pure $\mE$ field}
\label{pureE}

We first show the standard case: vacuum decay in a pure electric field. The imaginary part of action (Eq.\,(14) in~\cite{Labun:2012jf}) becomes
\begin{equation}
\label{ImaginarypureE}
\Im[\mL_\mathrm{EH}]=
\frac{m^2T^2}{8\pi}
\sum_{n=1}^\infty\frac{(-1)^n}{n^2}\cos[\frac g2 n\pi]e^{-nm/T}
\;.
\end{equation}
Comparing this result at $g=2.002319$ to that at $g=2$, the two are different by $<.1\%$ for $a<60 \mE_\mathrm{EH}$, and $<1\%$ for $60<a<100 \mE_\mathrm{EH}$.
 Thus pair production in a pure electric field is modified by an insignificant amount when correcting the electron's gyromagnetic ratio. We now move to vacuum instability in both $\mE$ and $\mB$ fields.

\subsection{Imaginary part of action: general form}
We compute the imaginary part of effective action for nonzero $a,b$. Using Eqs.\,(\ref{eq1}) and (\ref{g2}), which now produce a non-divergent action for the electron anomalous $g$-factor, we plug in a normalized expansion of the $\csch$ functions, typically used in computing scalar QED action~\cite{Cho:2000ei}:
\begin{align}
\mL_\mathrm{EH}=&\,
\frac{-1}{4\pi^3}\int_0^\infty \frac{ds}{s^3}e^{-im^2s} \cosh[\frac g2x] \cos[\frac g2y]
 \\[0.2cm] \nonumber
\hphantom=&
\!\!\!\!\!\!\!\!\!\!\!\!
\times
\sum_{n=1}^\infty\frac{(-1)^n}n
\left( x^3y\frac{\csch[n\pi y/x]}{x^2+n^2\pi^2}-xy^3\frac{\csch[n\pi x/y]}{y^2-n^2\pi^2}\right)
,
\end{align}
plus renormalizing terms, and where 
\begin{equation}
x=eas\;,\;y=ebs
\;.
\end{equation}
We can neglect the second term as it does not contribute to the imaginary part of action, and write the first part as:
\begin{align}
\mL_\mathrm{EH}=&\;
\frac{-1}{4\pi^3}\int_0^\infty \frac{ds}{s^3}(\cos[m^2s]-i\sin[m^2s]) 
 \\[0.2cm] \nonumber
\hphantom=&
\!\!\!\!\!\!\!\!\!\!\!\!
\times
\cosh[\frac g2x] \cos[\frac g2y]
\sum_{n=1}^\infty\frac{(-1)^n}n
x^3y\frac{\csch[n\pi b/a]}{x^2+n^2\pi^2}+\cdot\cdot\cdot
\;,
\end{align}
where the second term gives after summing residues
\begin{align}
\label{ImLEH}
\Im[\mL_\mathrm{EH}]=&\,
\frac{(ea)(eb)}{8\pi^2}\sum_{n=1}^\infty\frac{(-1)^n}n \cos[\frac g2n\pi] 
 \\[0.2cm] \nonumber
\hphantom=&
\hphantom{\frac{(ea)(eb)}{8\pi^2}}
\times
\frac{\cosh[\frac g2n\pi b/a]}{\sinh[n\pi b/a]}e^{-n\pi m^2/ea}
\;.
\end{align}
In the limit $b\to 0$ we recover Eq.\,(\ref{ImaginarypureE}):
\begin{align}
\Im[\mL_\mathrm{EH}]=&\,
\frac{e^2\mE^2}{8\pi^2}\sum_{n=1}^\infty\frac{(-1)^n}{n^2\pi} \cos[\frac g2n\pi] 
e^{-n\pi m^2/e\mE}
 \\[0.2cm] \nonumber
=&\,
\frac{m^2T^2}{8\pi}\sum_{n=1}^\infty\frac{(-1)^n}{n^2} \cos[\frac g2n\pi] 
e^{-nm/T}
\;,
\end{align}
the result for pure $\mE$ fields from~\cite{Labun:2012jf}.

\subsection{Evaluation and asymptotic behavior}

\begin{figure}[H]
\center
\includegraphics[width=1\columnwidth]{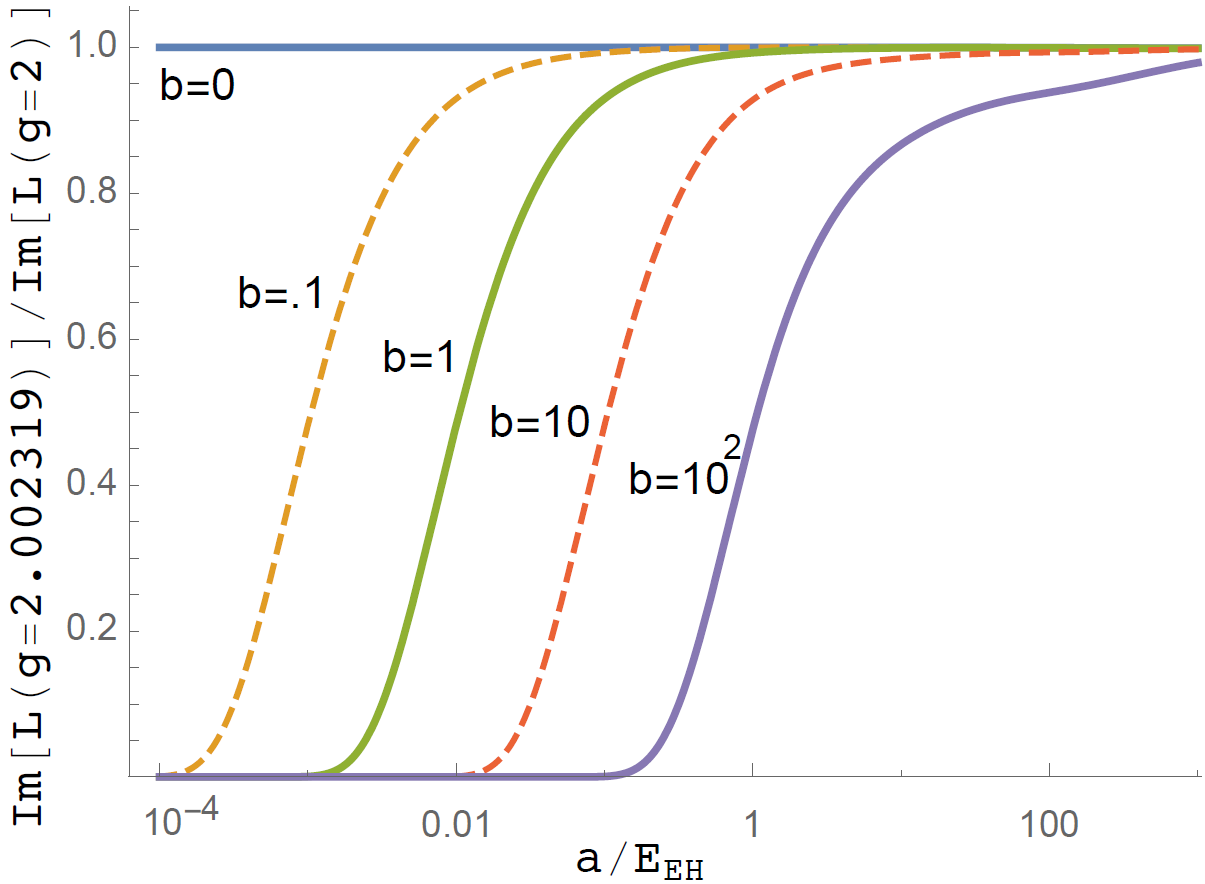}
\caption{\label{fig1} Ratio $\Im[\mL_\mathrm{EH}(a,b,g=2.002319)]/\Im[\mL_\mathrm{EH}(a,b,g=2)]$ is plotted against $a$, at fixed values of $b$ in units $\mE_\mathrm{EH}$.} 
\end{figure}

In contrast to the result for a pure $\mE$ field, section~\ref{pureE}, the result for arbitrary $a$ and $b$ obtained evaluating Eq.\,(\ref{ImLEH}) exhibits vacuum stabilization by the anomalous magnetic moment, see figure~\ref{fig1}, significant for large $b/a$. Pair production is completely suppressed as $b/a\to\infty$.

%
%
%
%

%
%
\begin{figure}[H]
\center
\includegraphics[width=1\columnwidth]{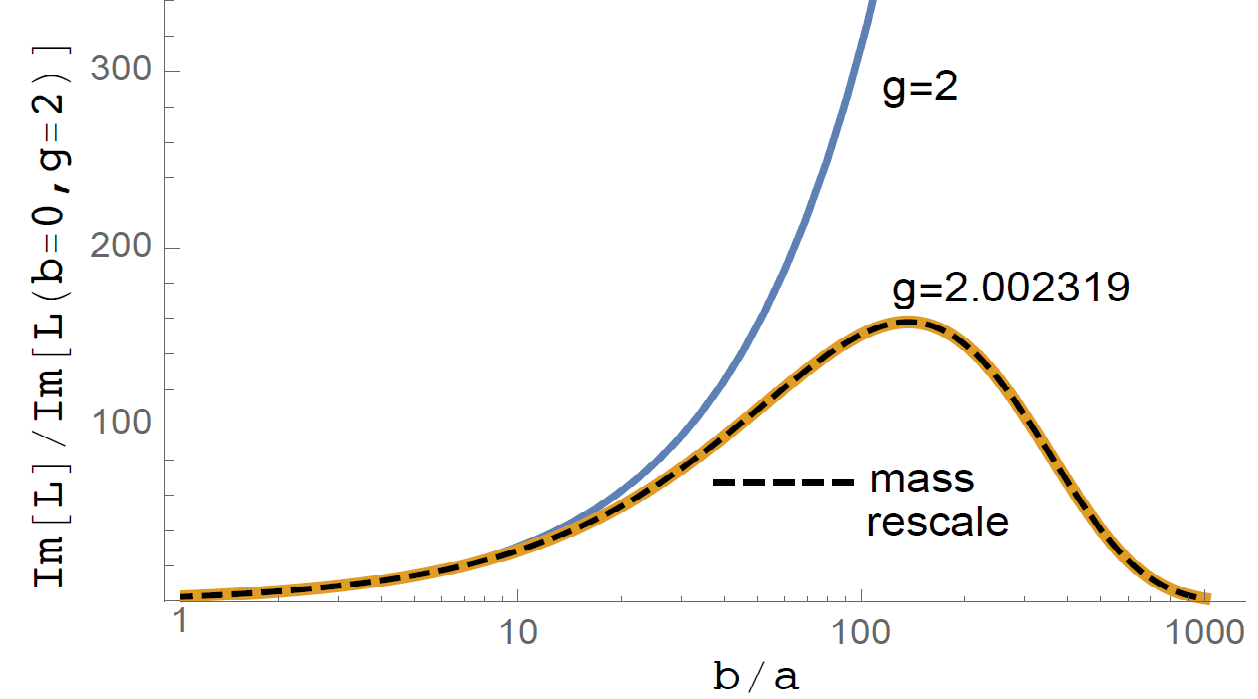}
\caption{\label{fig2} $\Im[\mL_\mathrm{EH}(a,b,g)]/\Im[\mL_\mathrm{EH}(a,b=0,g=2)]$ is depicted as a function $b/a$. Dashed line shows rescaled mass function from Eq.\,(\ref{rescaledmassUsed}) with $g=2.002319$.}
\end{figure}
%
%

\noindent Figure~\ref{fig2} offers the same result as in figure~\ref{fig1}, but with $\mB$ field dependence of pair production made more visible by taking the ratio with imaginary action for a pure $\mE$ field. We see that the anomalous $g$-induced suppression at large enough $b/a$ dominates the conventional $g=2$ vacuum decay enhancement by magnetic fields: the $e^+e^-$ production is maximized for $b/a\sim 10^2$. At $b/a\sim 10^3$, all effects cancel and we are left with a ratio of $\sim 1$ (pure $\mE$ field result). For fields of order $a\sim100\mE_\mathrm{EH}$ there is further suppression, lowering the peak in figure~\ref{fig2} ($\cos[\frac g2 n\pi]$ contribution discussed in section~\ref{pureE}).

To better understand the results shown in figure \ref{fig2}, we look at the asymptotic limit $b/a\to\infty$ in Eq.\,(\ref{ImLEH}):
\begin{align}
\label{rescaledmass}
\Im[\mL_\mathrm{EH}]\sim&\,
\frac{(ea)(eb)}{8\pi^2}\sum_{n=1}^\infty\frac{(-1)^n}n \cos[\frac g2n\pi] 
 \\[0.2cm] \nonumber
\hphantom=&
\hphantom{\frac{(ea)(eb)}{8\pi^2}}
\!\!\!\!\!\!\!\! \times
\left(1+2\sum_{r=1}^\infty e^{-2rn\pi b/a}\right)e^{-n\pi \widetilde m^2/ea}
\\[0.2cm] \label{rescaledmassUsed} 
\sim&\,
\frac{(ea)(eb)}{8\pi^2}\sum_{n=1}^\infty\frac{(-1)^n}n \cos[\frac g2n\pi]e^{-n\pi\widetilde m^2/ea}
\;,
\end{align}
where rescaled mass 
\begin{equation}
\label{tildem}
\widetilde m^2=m^2+\Big|\frac {|g_k-4k|}2-1\Big| b
\;,
\end{equation}
where we have written explicitly the transformation to $g_0$, Eq.\,(\ref{g2}), to demonstrate the periodicity as a function of $g$ of our result. The outer absolute value in Eq.\,(\ref{tildem}) assures that the mass always increases as is seen in  Eq.\,(\ref{rescaledmass}). The inner absolute value is due to evenness of the cos and cosh terms in  Eq.\,(\ref{ImLEH}). The result obtained using rescaled mass according to  Eq.\,(\ref{rescaledmassUsed}) is shown as a dashed line in figure \ref{fig2} and is valid for $b/a>1$. 

This form of mass rescaling shares similarities with mass catalysis computations~\cite{Gusynin:1995gt,Gusynin:1999ti,Ferrer:2000ed,Elizalde:2002ca,Shovkovy:2012zn}, since the leading correction to $g$ is linear in fine-structure constant $\alpha$.

\section{Conclusion}
An extension of the anomalous moment modification to two-loop effective action is of interest~\cite{Reuter:1996zm}. We have made here a step in this direction and have demonstrated that due to the electron's anomalous magnetic moment, vacuum instability by decay into $e^+e^-$ pairs experiences large modification for large $b/a$. This effect can be described in compact form by rescaling mass, Eq.\,(\ref{rescaledmassUsed}). We have shown that the effect of the gyromagnetic anomaly compensates and overwhelms the Schwinger $g=2$ result where $\mB$ enhances pair production.
We thus believe that an evaluation of electron-positron pair production for nearly constant fields on the scale of the order of electron Compton wave length $\lambda_C$, such that EH action is valid, must include this effect which dominates for strong $\mB$. We note that for field strengths $b/a\sim 10^3$, the vacuum decay is suppressed by factor $\sim \times10^{-3}$. In limit $b/a\to\infty$, pair production is completely removed. Our result is a step toward accounting for extreme magnetic field physics effects with $b/a\gg 1$ arising in heavy ion collisions and on magnetars. 

While not the focus of this work, we recognize a growing experimental interest in the high-intensity laser frontier~\cite{DiPiazza:2011tq,Hegelich:2014tda,Hegelich:2017iwb}. Seen our results, the modification of vacuum polarization for anomalous $g$-factors~\cite{Veltman:1997am,AngelesMartinez:2011nt,VaqueraAraujo:2012qa} should be reconsidered: our results differ due to recognition of the periodicity in $g$, potentially resulting in experimentally discernible differences between the cited results, and a periodic in $g$ result for the effect of vacuum polarization. Another concern we have is that in the prior work on the self-energy in leading~${\cal O}(\alpha)$, $\mB$-dependent corrections were found to be negative within the range $\mB<\mE_\mathrm{EH}$, while being positive for $\mB\gg\mE_\mathrm{EH}$~\cite{Jancovici:1970ep,Newton:1971pq,Constantinescu:1972qe,Tsai:1974id}. These results will need to be reinspected to account for the nonperturbative character of $g$-factor periodicity.

To conclude: we presented QED vacuum instability in strong magnetic fields allowing for $g\ne 2$. Our work is the first step towards a complete characterization at two loop order of strong $\mB$-field behavior, valid to all orders in magnetic field. We recognize new nonperturbative effects related to periodicity as function of $g$. Our results apply in straightforward fashion to consideration of $\mB$-dependence in $g\to g_f(a,b)$ and $m\to m_f(a,b)$. We have shown that pair production suppression occurs irrespective of the sign of the $g$-anomaly. We believe that the periodicity in $g$ exploited here will facilitate future computation of effective action including self-energy corrections.


\end{document}